\documentclass[print,twocolumn,amssymb,amsmath,prb,aps,showpacs,superscriptaddress]{revtex4-1}

\usepackage{graphicx}
\usepackage{dcolumn}
\usepackage{bm}
\usepackage{color}
\usepackage[dvips]{epsfig}
\usepackage[mathlines]{lineno}
\usepackage[colorlinks=true,linkcolor=blue]{hyperref}
\hypersetup{citecolor=blue}

\begin{document}

\title{Disorder suppressed charge-density-wave and its origin in 1\emph{T}-TaSe$_{2-x}$Te$_x$}

\author{Y. Liu}
\email[The authors contributed equally to this work.]{}
\affiliation{Key Laboratory of Materials Physics, Institute of Solid State Physics, Chinese Academy of Sciences, Hefei 230031, People's Republic of China}

\author{D. F. Shao}
\email[The authors contributed equally to this work.]{}
\affiliation{Key Laboratory of Materials Physics, Institute of Solid State Physics, Chinese Academy of Sciences, Hefei 230031, People's Republic of China}

\author{W. J. Lu}
\email[Corresponding author: ]{wjlu@issp.ac.cn}
\affiliation{Key Laboratory of Materials Physics, Institute of Solid State Physics, Chinese Academy of Sciences, Hefei 230031, People's Republic of China}

\author{L. J. Li}
\affiliation{Key Laboratory of Materials Physics, Institute of Solid State Physics, Chinese Academy of Sciences, Hefei 230031, People's Republic of China}

\author{H. Y. Lv}
\affiliation{Key Laboratory of Materials Physics, Institute of Solid State Physics, Chinese Academy of Sciences, Hefei 230031, People's Republic of China}

\author{X. D. Zhu}
\affiliation{High Magnetic Field Laboratory, Chinese Academy of Sciences, Hefei 230031, People's Republic of China}

\author{S. G. Tan}
\affiliation{Key Laboratory of Materials Physics, Institute of Solid State Physics, Chinese Academy of Sciences, Hefei 230031, People's Republic of China}

\author{B. Yuan}
\affiliation{Key Laboratory of Materials Physics, Institute of Solid State Physics, Chinese Academy of Sciences, Hefei 230031, People's Republic of China}

\author{L. Zu}
\affiliation{Key Laboratory of Materials Physics, Institute of Solid State Physics, Chinese Academy of Sciences, Hefei 230031, People's Republic of China}

\author{X. C. Kan}
\affiliation{Key Laboratory of Materials Physics, Institute of Solid State Physics, Chinese Academy of Sciences, Hefei 230031, People's Republic of China}

\author{W. H. Song}
\affiliation{Key Laboratory of Materials Physics, Institute of Solid State Physics, Chinese Academy of Sciences, Hefei 230031, People's Republic of China}

\author{Y. P. Sun}
\email[Corresponding author: ]{ypsun@issp.ac.cn}
\affiliation{High Magnetic Field Laboratory, Chinese Academy of Sciences, Hefei 230031, People's Republic of China}
\affiliation{Key Laboratory of Materials Physics, Institute of Solid State Physics, Chinese Academy of Sciences, Hefei 230031, People's Republic of China}
\affiliation{Collaborative Innovation Centre of Advanced Microstructures, Nanjing University, Nanjing 210093, People's Republic of China}

\begin{abstract}
In the sake of connecting the charge-density-wave (CDW) of TaSe$_2$ and single-\emph{\textbf{q}} CDW-type distortion of TaTe$_2$, we present an overall electronic phase diagram of 1\emph{T}-TaSe$_{2-x}$Te$_x$ ($0 \leq x \leq 2$). In the experimentally prepared single crystals, the CDW is completely suppressed as $0.5 < x < 1.5$, while superconductivity emerges as $0.2 < x < 1.2$. Theoretically, similar to 1\emph{T}-TaSe$_2$ and 1\emph{T}-TaTe$_2$, the hypothetic 1\emph{T}-TaSeTe with ordered Se/Ta/Te stacking shows instability in the phonon dispersion, indicating the presence of CDW in the ideally ordered sample. The contradictory between experimental and theoretical results suggests that the CDW is suppressed by disorder in 1\emph{T}-TaSe$_{2-x}$Te$_x$. The formation and suppression of CDW are found to be independent with Fermi surface nesting based on the generated electron susceptibility calculations. The calculation of phonon linewidth suggests the strong \textbf{\emph{q}}-dependent electron-phonon coupling induced period-lattice-distortion (PLD) should be related to our observation: The doping can largely distort the TaX$_6$ (X = Se, Te) octahedra, which are disorderly distributed. The resulted puckered Ta-Ta layers are not compatible with the two-dimensional PLD. Therefore, CDW is suppressed in 1\emph{T}-TaSe$_{2-x}$Te$_x$. Our results offer an indirect evidence that PLD, which can be influenced by strong disorder, is the origin of CDW in the system.
\end{abstract}


\maketitle

\section{Introduction}
The origin of charge-density-wave (CDW) is a rather old but still long-standing issue in condensed matter physics.\cite{Peierls-book-1955, Peierls-book-1991, Frohlich-1954} Since the discovery of CDW in transition-metal dichalcogenides (TMDs), the layered structures and various CDWs make the TMDs a model class of materials to investigate the mechanism of CDW.\cite{Wilson-CDW-review} Moreover, many typical TMDs show the coexistence and competition between superconductivity and CDW.\cite{Sipos-1T-TaS2-pressure, Morosan-CuxTiSe2, Morosan-PdxTiSe2, Wagner-CuxTaS2, LLJ-EPL, Ang-PRL, LY-APL, Ang-PRB, LY-JAP} The superconducting phase diagrams highly similar to those of unconventional superconductors  have been found.\cite{Morosan-CuxTiSe2, Morosan-PdxTiSe2, Wagner-CuxTaS2} Many experimental and theoretical works were performed to investigate the CDW in TMDs in the sake of figuring out the mechanism of CDW and the interplay between CDW and superconductivity.\cite{Weber-PRL, Dai-PRB, Johannes-nesting, AmyLiu-1T-TaS2, AmyLiu-1T-TaSe2} However, the origin of CDW is still under debate.

The CDW and accompanied period-lattice-distortion (PLD) are usually explained by Peierls picture.\cite{Peierls-book-1955, Peierls-book-1991, Frohlich-1954, Johannes-nesting} In this picture, Fermi surface nesting, a pure electronic effect, drives the charge redistribution regardless of whether or not PLD subsequently happens. There is an opposite mechanism that the charge redistribution is driven by strong \emph{\textbf{q}}-dependent electron-phonon coupling induced PLD, while Fermi surface nesting only plays a minor role.\cite{Chan-1973} Johannes \emph{et al.}\cite{Johannes-nesting} concluded that if a material with CDW originates from Fermi surface nesting, the generated electron susceptibility ($\chi$) must show peaks of its real part ($\chi^{\prime}$) and imaginary part ($\chi^{\prime\prime}$) at the CDW vector ($\emph{\textbf{q}}_{\textrm{CDW}}$), and all phonons must soften at the same vector (although in theory it is unavoidable,\cite{Chiba-2003} such softening for multiple modes has never been actually observed in real materials. For example, ZrTe$_3$\cite{Hoesch-2009} and KCP\cite{Renker-1973} have strong nesting at the CDW ordering wave vectors but only one mode softens). According to this conclusion, it seems to be clear that the Fermi surface nesting is ruled out in the origin of CDW in 2\emph{H}-TMDs: The density-functional-theory (DFT) calculations show that there are no peaks of $\chi^{\prime\prime}$ found at $\emph{\textbf{q}}_{\textrm{CDW}}$ for 2\emph{H}-NbSe2 and 2\emph{H}-TaSe2, while $\chi^{\prime}$ shows weak peaks at $\emph{\textbf{q}}_{\textrm{CDW}}$.\cite{Johannes-nesting, Johannes-NbSe2,Calandra-NbSe2} The PLD mechanism in 2\emph{H}-TMDs is recently supported by more and more theoretical and experimental studies.\cite{AmyLiu-2H-TaSe2, Weber-PRL, Dai-PRB}

However, in 1\emph{T}-TMDs, the situation seems to be complex. Both 1\emph{T}-TaS$_2$ and 1\emph{T}-TaSe$_2$ show 13.9$^\circ$ rotated  $\sqrt{13}\times\sqrt{13}$ commensurate-CDW (CCDW).\cite{Wilson-CDW-review} The early reports show the calculated Fermi surface nesting vectors are well corresponding to $\emph{\textbf{q}}_{\textrm{CCDW}}=\frac{3}{13}\emph{\textbf{a}}^*+\frac{1}{13}\emph{\textbf{b}}^*$,\cite{Wilson-CDW-review, TaS2-xray, Myron-1975, Myron-1977} while the recent DFT calculations do not reflect the nesting vector at $\emph{\textbf{q}}_{\textrm{CCDW}}$.\cite{AmyLiu-1T-TaS2, AmyLiu-1T-TaSe2, Yu-arxiv} Strong electron-phonon coupling strengths at $\emph{\textbf{q}}_{\textrm{CDW}}$ are obtained from the density-functional-perturbation-theory (DFPT)calculations for 1\emph{T}-TaS$_2$ and 1\emph{T}-TaSe$_2$ under pressure,\cite{DFPT,AmyLiu-1T-TaS2, AmyLiu-1T-TaSe2} which supports the PLD mechanism. On the other hand, NbTe$_2$ and TaTe$_2$ are typical TMDs with monoclinic distorted-1\emph{T} structure, for which the distortion could be considered as single-\emph{\textbf{q}} CDW-type distortion with a vector of $\emph{\textbf{q}}=\frac{1}{3}\emph{\textbf{a}}^*$.\cite{Wilson-CDW-review, Battaglia-NbTe2, Sharma-TaX2} Battaglia \emph{et al}. suggested that the Fermi surface nesting plays a key role in the formation of such single-\emph{\textbf{q}} CDW.\cite{Battaglia-NbTe2} A convincing origin of CDW in 1\emph{T}-TMDs could not be obtained from the above confusing and contradictory results so far. More experimental evidences and theoretical investigations are needed to figure out the puzzle.

Previously, we reported the electronic phase diagrams for 1\emph{T}-TaS$_{2-x}$Se$_x$ ($0 \leq x \leq 2$) and $4H_\textrm{b}$-TaS$_{2-x}$Se$_x$ ($0 \leq x \leq 1.5$), in which the isovalent substitution does not completely suppress CDW of the end members.\cite{LY-APL,LY-JAP} If similar characteristic still exists in 1\emph{T}-TaSe$_{2-x}$Te$_x$ ($0 \leq x \leq 2$), a potential gradual variation from $\emph{\textbf{q}}_{\textrm{CDW}}$ of 1\emph{T}-TaSe$_2$ to that of 1\emph{T}-TaTe$_2$ might be observed. Therefore, in the present work, we prepared a series of 1\emph{T}-TaSe$_{2-x}$Te$_x$ ($0 \leq x \leq 2$) single crystals via the chemical-vapor-transport (CVT) method and obtained an overall electronic phase diagram through the transport measurements. Surprisingly, the CDW disappears in 1\emph{T}-TaSe$_{2-x}$Te$_x$ as $0.5 < x < 1.5$, which is beyond our expectation and destroys the connection of 1\emph{T}-TaSe$_2$ and 1\emph{T}-TaTe$_2$. Superconductivity emerges when $0.2 < x < 1.2$. According to phonon calculations, the hypothetic ordered 1\emph{T}-TaSeTe shows phonon instability similar to 1\emph{T}-TaSe$_2$ and 1\emph{T}-TaTe$_2$, indicating the CDW should exist if the doped sample is ideally ordered. It means that the disorder in experimentally prepared samples suppresses CDW. The formation and suppression of CDW are found to be independent with Fermi surface nesting based on the generated electron susceptibility calculations. The calculation of phonon linewidth suggests the CDW instability is strongly coupled with the \textbf{\emph{q}}-dependent electron-phonon coupling induced period-lattice-distortion (PLD). Through analysis of the optimized structures of 1\emph{T}-TaSe$_2$, 1\emph{T}-TaTe$_2$, and 1\emph{T}-TaSeTe with ordered Se/Ta/Te stacking structure, we found that the doping can largely distort the TaX$_6$ (X = Se, Te) octahedra. The disordered distribution of such distorted octahedra will pucker Ta-Ta layers, which is not compatible with the two-dimensional PLD. That might be the reason why the CDW is suppressed in 1\emph{T}-TaSe$_{2-x}$Te$_x$. The interplay beween CDW and superconductivity is also discussed. Our results offer an indirect evidence that the CDW originates from PLD, which can be influenced by strong disorder.

\section{Experiment and calculation details}
The single crystals were grown via the CVT method with iodine as a transport agent. The high-purity elements Ta, Se, and Te were mixed in chemical stoichiometry, and heated at 900$^\circ$C for 4 days in an evacuated quartz tube. The harvested TaSe$_{2-x}$Te$_x$ powders and iodine (density: 5 mg/cm$^3$) were then sealed in an another quartz tube and heated for two weeks in a two-zone furnace, where the temperature of source and growth zones were fixed at 950$^\circ$C and 850$^\circ$C, respectively. The tubes were rapidly quenched in cold water to ensure retaining of the 1\emph{T} phase. The X-ray diffraction (XRD) patterns were obtained on a Philips X$^\prime$pert PRO diffractometer with Cu $K_\alpha$ radiation ($\lambda$ = 1.5418 {\AA}). Structural refinements were performed by using Rietveld method with the X$^\prime$pert HighScore Plus software. The average stoichiometry was determined by examination of multiple points using X-ray energy dispersive spectroscopy (EDS) with a scanning electron microscopy (SEM). The EDS results indicate that the actual concentration \emph{x} is close to the nominal one. The resistivity ($\rho$) measurements down to 2.0 K were carried out by the standard four-probe method in a Quantum Design Physical Property Measurement System (PPMS).

The DFT calculations were carried out using QUANTUM ESPRESSO package with ultrasoft pseudopotentials.\cite{QE} The exchange-correlation interaction was treated with the local-density-approximation (LDA) according to Perdew and Zunger.\cite{pz} The energy cutoff for the plane-wave basis set was 35 Ry. Brillouin zone sampling is performed on the Monkhorst-Pack (MP) mesh of $16 \times 16 \times 8$.\cite{MP} The Vanderbilt-Marzari Fermi smearing method with a smearing parameter of $\sigma=0.02$ Ry was used for the calculations of the total energy and electron charge density. Phonon dispersions were calculated using DFPT with an $8 \times 8 \times 4$ mesh of \emph{\textbf{q}}-points. In order to investigate the electron-phonon coupling around the \textbf{\emph{q}}$_\textrm{CDW}$, denser $64 \times 64 \times 8$ and $32 \times 32 \times 4$ mesh of \textbf{\emph{k}}-points, $16 \times 16 \times 1$ \textbf{\emph{q}}-points were used.

\section{Results and discussion}

\begin{figure}
\includegraphics[width=0.94\columnwidth]{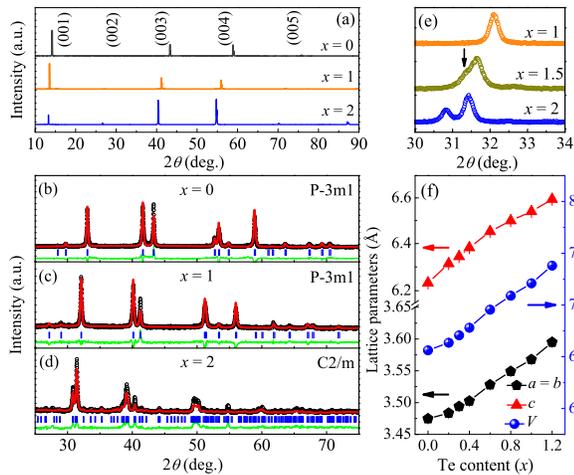}
\caption{(Color online) (a) Single-crystal XRD patterns of 1\emph{T}-TaSe$_{2-x}$Te$_x$ for \emph{x} = 0, 1, and 2. (b)-(d) Powder XRD patterns with Rietveld refinements of 1\emph{T}-TaSe$_{2-x}$Te$_x$ for \emph{x} = 0, 1, and 2, respectively. (e) The enlargement of the (011) peaks of the powder XRD patterns of 1\emph{T}-TaSe$_{2-x}$Te$_x$ for \emph{x} = 1, 1.5, and 2. (f) Evolution of lattice parameters (\emph{a}, \emph{c}) and cell volume (\emph{V}) of 1\emph{T}-TaSe$_{2-x}$Te$_x$.}
\label{Fig_XRD}
\end{figure}

The single crystal XRD patterns of 1\emph{T}-TaSe$_{2-x}$Te$_x$ ($x$ = 0, 1, and 2) are shown in Fig.~\ref{Fig_XRD}(a), in which only ($00l$) reflections were observed, suggesting the \emph{c}-axis is perpendicular to the surface of single crystal. With increasing \emph{x}, the diffraction peaks distinctly shift to lower angles, reflecting the crystal expansion induced by Te doping. To further confirm the structure, several single crystals were crushed and used in the powder XRD experiment. Figures~\ref{Fig_XRD}(b)-(d) show the powder XRD patterns and the structural refinement results of Rietveld analysis for the selected samples with \emph{x} = 0, 1, and 2. Figure~\ref{Fig_XRD}(e) shows the enlargement of the (011) peak for \emph{x} = 1, 1.5, and 2. It shows that the ideal CdI$_2$-type 1\emph{T} structure for \emph{x} = 1 leads to a single (011) peak while there are double peaks in the monoclinic distorted-1\emph{T} structure for \emph{x} = 2. One should notice that the (011) peak starts to split when \emph{x} = 1.5, indicating the emergence of distorted-1\emph{T} structure, as shown in Fig.~\ref{Fig_XRD}(e). The evolution of lattice parameters (\emph{a}, \emph{c}) and unit cell volume (\emph{V}) of 1\emph{T}-TaSe$_{2-x}$Te$_x$ are depicted in Fig.~\ref{Fig_XRD}(f). Indeed, the values of \emph{a}, \emph{c}, and \emph{V} monotonously increase with \emph{x}, in accordance with the larger ion radius of Te than that of Se.

\begin{figure}
\includegraphics[width=0.9\columnwidth]{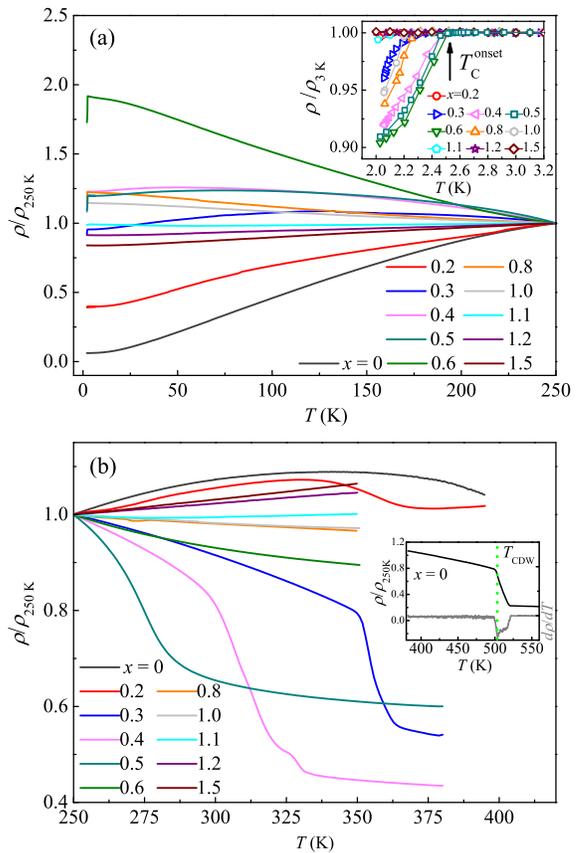}
\caption{(Color online) Temperature dependence of in-plane resistivity ratio ($\rho/\rho_{\textrm{250K}}$) of 1\emph{T}-TaSe$_{2-x}$Te$_x$ below \emph{T} = 250 K (a) and above \emph{T} = 250 K (b). The insets are the enlargement of superconducting transitions at low temperatures and the CCDW transition of 1\emph{T}-TaSe$_2$ at high temperatures.}
\label{Fig_resistivity}
\end{figure}

\begin{figure}
\includegraphics[width=0.9\columnwidth]{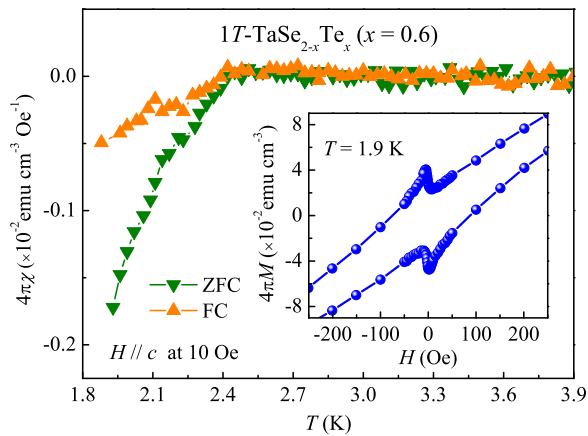}
\caption{(Color online) Temperature dependence of magnetic susceptibility ($4\pi\chi$) for $x = 0.6$. Inset: the magnetization hysteresis loop obtained at \emph{T} = 1.9 K with magnetic field \emph{H} paralleling \emph{c}-axis.}
\label{Fig_magnetic}
\end{figure}

Figure~\ref{Fig_resistivity} shows the temperature dependences of in-plane resistivity ratio ($\rho/\rho_{\textrm{250K}}$) of 1\emph{T}-TaSe$_{2-x}$Te$_x$ single crystals. The room temperature resistivity is about 1.4 m$\Omega$ cm for \emph{x} = 0. The values of resistivity for the Te-doped samples keep at this order of magnitude. In addition, the residual resistivity ratio (RRR = $\rho_{\textrm{300K}}/\rho_{\textrm{3K}}$) is calculated as 17.4 for $x = 0$. For the Te-doped samples, it sharply decreases to 2.66 for $x = 0.2$, and reaches the minimum 0.48 for $x = 0.6$, and then slightly increases to 1.13 for $x = 1.5$ (the value of RRR is 0.85 for $x = 1$). The small values of RRR may be as a reflection of the substantial disorder present in Te-doped samples, which will be further discussed in the calculation part. As shown in the inset of Fig.~\ref{Fig_resistivity}(a), the signature of superconductivity emerges as $x \geq 0.2$, and finally disappears for $x \geq 1.2$ within our measurement limitation $T \geq 2$ K. The maximum of superconducting onset temperature ($T_\textrm{c}^{\textrm{onset}}$) is 2.5 K for \emph{x} = 0.6. To further confirm the superconductivity, we also measured the magnetic properties of the superconducting samples at \emph{H} = 10 Oe with the magnetic field paralleling to the \emph{c}-axis. As an example, Fig.~\ref{Fig_magnetic} shows the result of the optimal sample 1\emph{T}-TaSe$_{1.4}$Te$_{0.6}$. Undoubtedly, the diamagnetism signal at low temperatures demonstrates the occurrence of superconductivity, of which the transition temperature (\emph{T}$_\textrm{c}$) defined by the onset point of zero-field-cooling (ZFC) and field-cooling (FC) curves is 2.45 K for $x = 0.6$. The inset of Fig.~\ref{Fig_magnetic} shows the magnetization hysteresis loop \emph{M}(\emph{H}) obtained at \emph{T} = 1.9 K, which shows it is a typical type-II superconductor.

Figure~\ref{Fig_resistivity}(b) shows the high temperature part of $\rho/\rho_{\textrm{250K}}$. As shown in the inset of Fig.~\ref{Fig_resistivity}(b), 1\emph{T}-TaSe$_2$ exhibits a CDW transition at $\emph{T}_{\textrm{CDW}} \sim 503$ K, which is defined by the minimum of $d\rho/dT$, with the formation of $\sqrt{13}\times\sqrt{13}$ superstructure. The resistivity shows an upturn upon cooling due to the presence of gapping in the Fermi surface. With increasing Te content, the CDW transition gradually shifts to lower temperatures. It can be found that the CDW unexpectedly disappears when $x > 0.5$.

\begin{figure}
\includegraphics[width=0.9\columnwidth]{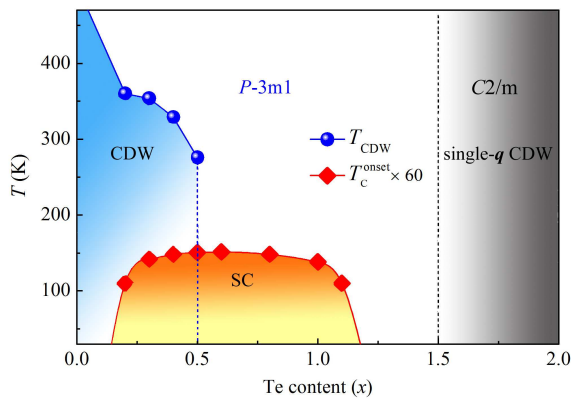}
\caption{(Color online) Electronic phase diagram of 1\emph{T}-TaSe$_{2-x}$Te$_x$ as a function of temperature and Te content.}
\label{Fig_phase}
\end{figure}

Figure~\ref{Fig_phase} summarizes the overall electronic phase diagram as a function of temperature and doping level in 1\emph{T}-TaSe$_{2-x}$Te$_x$. The \emph{x} dependence of $T_\textrm{c}^{\textrm{onset}}$ follows a dome-like shape. The superconductivity induced by isovalent doping is similar to that in 1\emph{T}-TaS$_{2-x}$Se$_x$.\cite{LY-APL} However, the CDW is gradually suppressed by Te doping and disappears as $x > 0.5$, which is quite different from the situation in 1\emph{T}-TaS$_{2-x}$Se$_x$.\cite{LY-APL} With heavier Te content $x > 1.5$, the crystal structure gradually distorts to monoclinic distorted-1\emph{T} structure, which could also be considered as a single-\emph{\textbf{q}} CDW-type distortion.\cite{Wilson-CDW-review, Sharma-TaX2}

To investigate the suppression of CDW in 1\emph{T}-TaSe$_{2-x}$Te$_x$ system, we calculated the two end members of 1\emph{T}-TaSe$_2$ and 1\emph{T}-TaTe$_2$, and the simplest hypothetical sample 1\emph{T}-TaSeTe with an ordered stacking of Se/Ta/Te, which is represented as 1\emph{T}-TaSeTe(O). The fully optimized structural parameters, as listed in Table~\ref{Structure_LDA}, are close to those from the previous LDA calculation.\cite{AmyLiu-1T-TaSe2} The underestimation of lattice parameters is expectable for LDA.\cite{AmyLiu-1T-TaSe2}

\begin{table}
\caption{\label{Structure_LDA}Structural parameters fully optimized by LDA for 1\emph{T}-TaSe$_2$, 1\emph{T}-TaSeTe(O),and 1\emph{T}-TaTe$_2$.}
\begin{ruledtabular}
\begin{tabular}{cccc}
&\emph{a}({\AA}) &\emph{c}({\AA}) &$z_{\textrm{X}}$ \\
\hline TaSe$_2$ & 3.406 & 6.086 & $z_{\textrm{Se}}=\pm0.271$ \\
TaSeTe(O) & 3.507 & 6.337 & $z_{\textrm{Se}}=0.249$,$z_{\textrm{Te}}=-0.296$ \\
TaTe$_2$ & 3.622 & 6.572 & $z_{\textrm{Te}}=\pm0.274$ \\
\end{tabular}
\end{ruledtabular}
\end{table}

\begin{figure*}
\includegraphics[width=2\columnwidth]{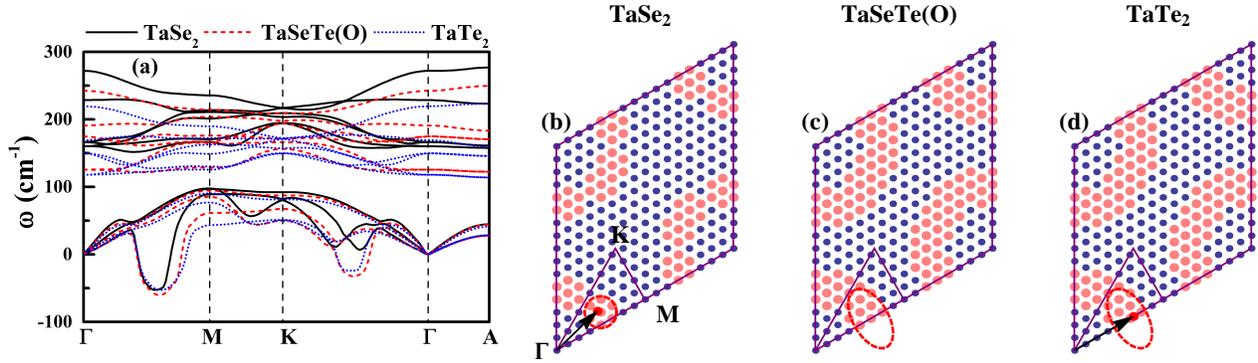}
\caption{(Color online) (a) Phonon dispersions of 1\emph{T}-TaSe$_2$, 1\emph{T}-TaSeTe(O), and 1\emph{T}-TaTe$_2$. (b), (c), and (d) show the calculated $16 \times 16 \times 1$ \textbf{\emph{q}}-points in the $\emph{\textbf{q}}_z = 0$ plane for 1\emph{T}-TaSe$_2$, 1\emph{T}-TaSeTe(O), and 1\emph{T}-TaTe$_2$, respectively. The pink solid circles in (b)-(d) denote the \textbf{\emph{q}}-points in which the frequency of the lowest mode is imaginary. The red dashed circle and ellipses denote the areas of phonon instability. The red solid circles and the black arrows in (b) and (d) denote the reported $\emph{\textbf{q}}_{\textrm{CCDW}}=\frac{3}{13}\emph{\textbf{a}}^*\textrm{+}\frac{1}{13}\emph{\textbf{b}}^*$ and $\emph{\textbf{q}}\approx\frac{1}{3}\emph{\textbf{a}}^*$ for 1\emph{T}-TaSe$_2$ and 1\emph{T}-TaTe$_2$, respectively. The high-symmetry points are shown in (b).}
\label{Fig_phonon}
\end{figure*}

\begin{figure*}
\includegraphics[width=1.8\columnwidth]{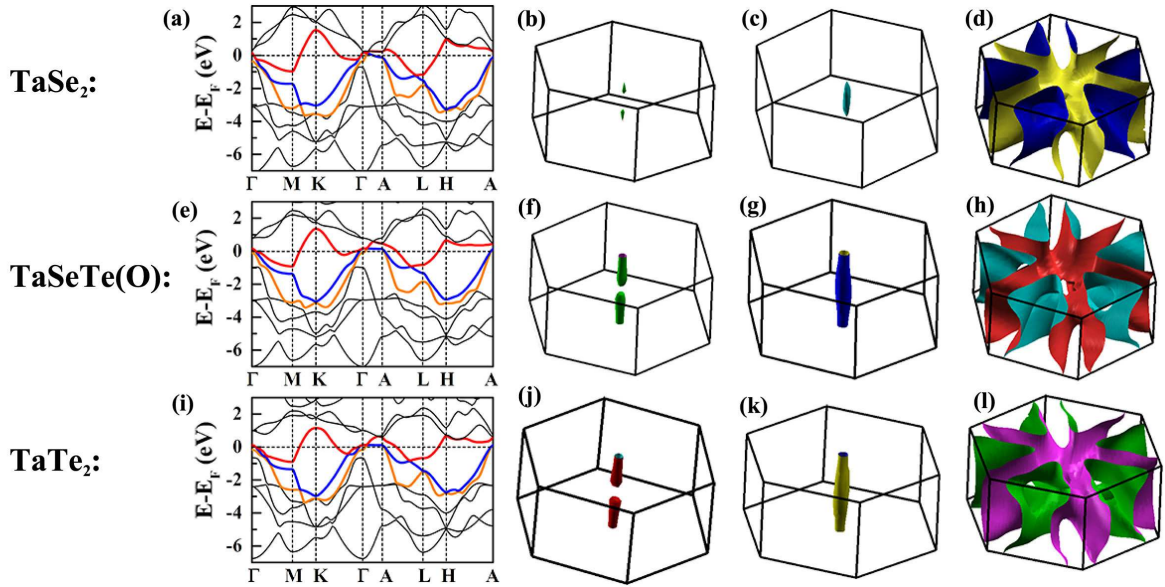}
\caption{(Color online) Top panel: Band structure (a) and Fermi surfaces (b), (c), and (d) of 1\emph{T}-TaSe$_2$. Middle panel: Band structure (e) and Fermi surfaces (f), (g), and (h) of a hypothetic ordered 1\emph{T}-TaSeTe. Bottom panel: Band structure (i) and Fermi surfaces (j), (k), and (l) of a hypothetic 1\emph{T}-TaTe$_2$. The bands crossing the $E_\textrm{F}$ are colored in red, blue, and orange, respectively.}
\label{Fig_band}
\end{figure*}

\begin{figure*}
\includegraphics[width=1.8\columnwidth]{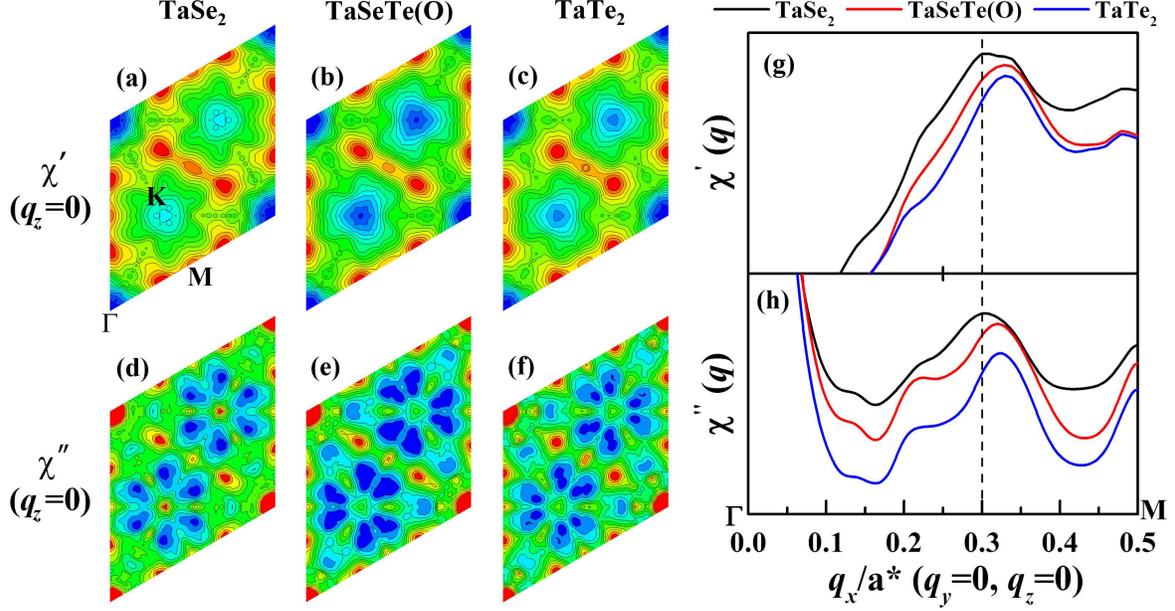}
\caption{(Color online) Upper left panel: The cross section of real part of the generated electron susceptibility with $\emph{\textbf{q}}_z$ = 0 for 1\emph{T}-TaSe$_2$ (a), 1\emph{T}-TaSeTe(O) (b), and 1\emph{T}-TaTe$_2$ (c). Bottom left panel: The cross section of imaginary part of the generated electron susceptibility with $\emph{\textbf{q}}_z$ = 0 for 1\emph{T}-TaSe$_2$, 1\emph{T}-TaSeTe(O) (f), and 1\emph{T}-TaTe$_2$ (g). Right panel shows the real and imaginary parts of the generated electron susceptibilities along the reciprocal unit cell boundary. In these contour maps of (a)-(f), the highest and lowest values are denoted as color in red and blue, respectively. The dashed line in right panel denotes the location of the peak of the real and imaginary part of the generated electron susceptibility for 1\emph{T}-TaSe$_2$ at $\emph{\textbf{q}}=0.3\emph{\textbf{a}}^*$. The high-symmetry points are denoted in (a).}
\label{Fig_nesting}
\end{figure*}

\begin{figure}
\includegraphics[width=0.98\columnwidth]{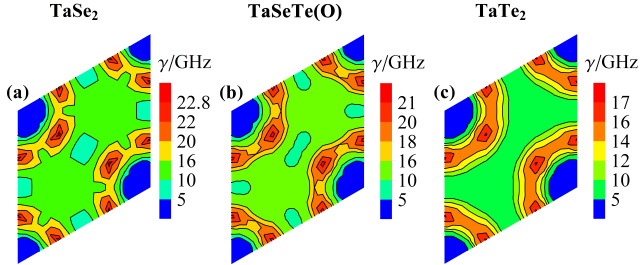}
\caption{(Color online) The contour map of the phonon linewidth $\gamma$ of the lowest phonon modes in the $\emph{\textbf{q}}_z$ = 0 plane for 1\emph{T}-TaSe$_2$ (a), 1\emph{T}-TaSeTe(O) (b), and 1\emph{T}-TaTe$_2$ (c).}
\label{Fig_gamma}
\end{figure}

\begin{figure}
\includegraphics[width=0.9\columnwidth]{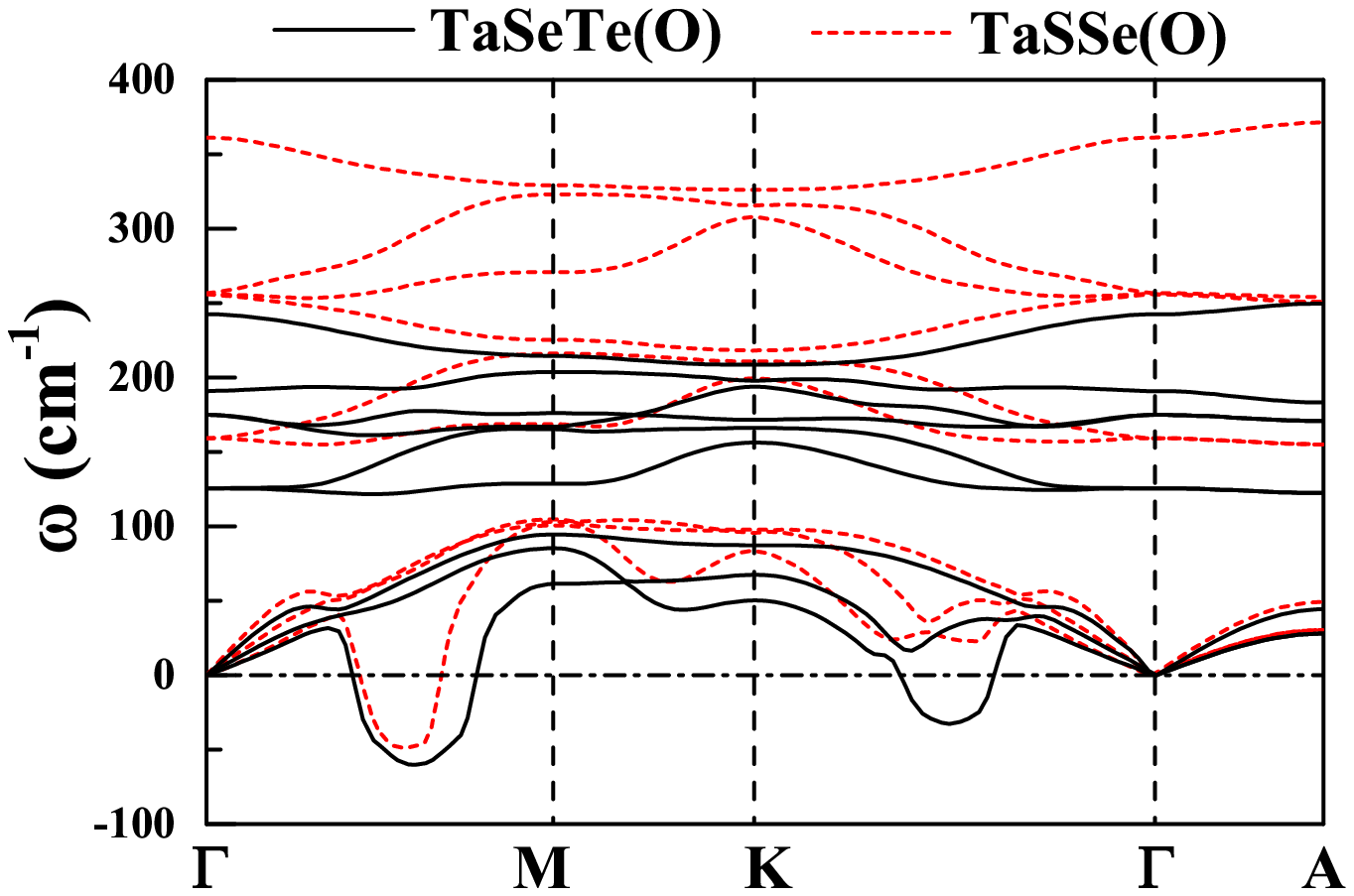}
\caption{(Color online) Phonon dispersions of 1\emph{T}-TaSeTe(O) and 1\emph{T}-TaSSe(O).}
\label{Fig_TaSSe_phonon}
\end{figure}

The phonon calculation is an effective method to simulate the CDW instability: The calculated phonon instabilities of 2\emph{H}-NbSe$_2$,\cite{Johannes-NbSe2} 2\emph{H}-TaSe$_2$,\cite{AmyLiu-2H-TaSe2} 1\emph{T}-NbTe$_2$,\cite{Battaglia-NbTe2} and 1\emph{T}-TaX$_2$ (X = S, Se, Te),\cite{AmyLiu-1T-TaS2, AmyLiu-1T-TaSe2, Battaglia-NbTe2} just locate at the CDW vectors of these TMDs. Moreover, Liu \emph{et al.} show the calculated phonon instability of 1\emph{T}-TaS$_2$ is suppressed by pressure while the disappearance of CDW for pressurized 1\emph{T}-TaS$_2$ is experimentally confirmed,\cite{AmyLiu-1T-TaS2,Sipos-1T-TaS2-pressure} conclusively proving that the calculated phonon instability can correctly reflect the CDW instability. Therefore, our theoretical investigation of 1\emph{T}-TaSe$_{2-x}$Te$_x$ started with the phonon calculation. Figure~\ref{Fig_phonon}(a) shows the phonon dispersions of 1\emph{T}-TaSe$_2$, 1\emph{T}-TaSeTe(O), and 1\emph{T}-TaTe$_2$. For 1\emph{T}-TaSe$_2$, the phonon dispersions are in good agreement with the previous calculation by Ge \emph{et al.}.\cite{AmyLiu-1T-TaSe2} As shown in Fig.~\ref{Fig_phonon}(b), one can notice that the calculated instability is just around the reported CCDW vector ($\emph{\textbf{q}}_{\textrm{CCDW}}=\frac{3}{13}\emph{\textbf{a}}^*\textrm{+}\frac{1}{13}\emph{\textbf{b}}^*$).\cite{TaS2-xray} For 1\emph{T}-TaTe$_2$, the area of instability is centered near $\emph{\textbf{q}}\approx\frac{1}{3}\emph{\textbf{a}}^*$, which is corresponding to the reported ($3\times1$) single-\emph{\textbf{q}} CDW-type superlattice.\cite{Battaglia-NbTe2} The high coincidence of the calculated and experimentally reported instabilities strongly proves the reliability and accuracy of the phonon calculation in the CDW systems. Moreover, different from 1\emph{T}-TaSe$_2$, 1\emph{T}-TaTe$_2$ shows much larger area of instability, which expands to the $\Gamma$K line. That might be the reason why 1\emph{T}-TaSe$_2$ only shows small atomic displacement in CDW phase, which can be suppressed by high temperature, while the single-\textbf{\emph{q}} CDW-type distortion in TaTe$_2$ is very stable and the ideal 1\emph{T} structure has never been observed.

Surprisingly, for 1\emph{T}-TaSeTe(O), a phonon instability can be found as well, as shown in Figs.~\ref{Fig_phonon}(a) and (c). The area of instability is very similar to that of 1\emph{T}-TaTe$_2$, implying that if 1\emph{T}-TaSeTe with such ordered structure exists, CDW (single-\textbf{\emph{q}} type) should be observed as in the case of 1\emph{T}-TaTe$_2$. However, according to the experimentally obtained phase diagram of Fig.~\ref{Fig_phase}, one should notice that the CDW or CDW-type distortion only exists in the Se-rich or Te-rich areas (the concentration $\geq75\%$), in which the disorder is not prominent. In the middle area, stronger disorder can be expected, which might be responsible for the disappearance of CDW.

In order to figure out how the disorder influences the CDW, we further performed some indirect calculations to investigate whether the disorder affects the Fermi surface nesting or PLD. Figure~\ref{Fig_band} shows the band structures and Fermi surfaces of 1\emph{T}-TaSe$_2$, 1\emph{T}-TaSeTe(O), and 1\emph{T}-TaTe$_2$. For 1\emph{T}-TaSe$_2$, the early calculations show there is only one band crossing Fermi energy ($E_\textrm{F}$), which does not cross $E_\textrm{F}$ in the vicinity of $\Gamma$ point.\cite{Myron-1975, Woolley-1977} Moreover, there is a gap (about 0.1 $\sim$ 0.2 eV) below the band crossing $E_\textrm{F}$.\cite{Myron-1975, Woolley-1977} However, the recent angle-resolved photoemission (ARPES) experiment clearly shows a hybridization of bands at $\Gamma$ close to $E_\textrm{F}$, where small hole-type pockets are observed.\cite{Bovet-ARPES} Obviously, our LDA calculations accurately simulated the band structure of 1\emph{T}-TaSe$_2$ (Fig.~\ref{Fig_band}(a)). Three bands cross $E_\textrm{F}$: The lower two bands (colored in blue and orange in Fig.~\ref{Fig_band}(a)) form small cylindrical hole-type pockets close to $\Gamma$ (Figs.~\ref{Fig_band}(b) and (c)); The Fermi surface introduced from the higher band crossing $E_\textrm{F}$ (colored in red in Fig.~\ref{Fig_band}(a)) is shown in Fig.~\ref{Fig_band}(d). For 1\emph{T}-TaSeTe(O) and 1\emph{T}-TaTe$_2$, one can notice that the band structures and Fermi surfaces (Figs.~\ref{Fig_band}(e)-(l)) are highly similar to those of 1\emph{T}-TaSe$_2$.

The Fermi surface nesting can be reflected in generated electron susceptibility. The real part of the electron susceptibility is defined as
\begin{equation}
\chi^{\prime}(\emph{\textbf{q}})=\sum_\emph{\textbf{k}}\frac{f(e_\emph{\textbf{k}})-f(e_{\emph{\textbf{k}}\textrm{+}\emph{\textbf{q}}})}{e_\emph{\textbf{k}}-e_{\emph{\textbf{k}}\textrm{+}\emph{\textbf{q}}}}, \end{equation}
where $f(e_\textbf{\emph{k}})$ is Fermi-Dirac function. The imaginary part is\cite{Johannes-nesting}
\begin{equation}
\chi^{\prime\prime}(\emph{\textbf{q}})=\sum_\emph{\textbf{k}}\delta(e_\emph{\textbf{k}}-e_\emph{\textbf{F}})\delta(e_{\emph{\textbf{k}}\textrm{+}\emph{\textbf{q}}}-e_\emph{\textbf{F}}).
\end{equation}
We used a mesh of approximately 40,000 \emph{\textbf{k}} points in the full reciprocal unit cell to calculate the energy eigenvalues derived for the electron susceptibilities. Figures~\ref{Fig_nesting}(a)-(f) show the cross sections of $\chi^{\prime}$ and $\chi^{\prime\prime}$ with $\emph{\textbf{q}}_z$ = 0 for 1\emph{T}-TaSe$_2$, 1\emph{T}-TaSeTe(O), and 1\emph{T}-TaTe$_2$. We found that all the maxima of $\chi^{\prime}$ and $\chi^{\prime\prime}$ of 1\emph{T}-TaSe$_2$, 1\emph{T}-TaSeTe(O), and 1\emph{T}-TaTe$_2$ locate between $\Gamma$ and M points. And the maxima shift towards M point with increasing Te content (Figures~\ref{Fig_nesting}(g) and (h)). For 1\emph{T}-TaSe$_2$, both the maxima of $\chi^{\prime}$ and $\chi^{\prime\prime}$ locate at $\emph{\textbf{q}}\approx0.3\emph{\textbf{a}}^*$. Earlier calculation by Myron \emph{et al}. shows a peak of $\chi^{\prime}$ at $\emph{\textbf{q}}\approx0.28\emph{\textbf{a}}^*$,\cite{Myron-1977} while recent calculation by Yu \emph{et al}. reports a maximum of $\chi^{\prime}$ at $\emph{\textbf{q}}\approx0.295\emph{\textbf{a}}^*$.\cite{Yu-arxiv} Clearly, the maxima of $\chi^{\prime}$ and $\chi^{\prime\prime}$ locate far away from $\emph{\textbf{q}}_{\textrm{CCDW}}=\frac{3}{13}\emph{\textbf{a}}^*+\frac{1}{13}\emph{\textbf{b}}^*$. Although Myron \emph{et al}. suggested that the peak of $\chi^{\prime}$ at $\emph{\textbf{q}} = 0.25\emph{\textbf{a}}^* \sim 0.30\emph{\textbf{a}}^*$ can lead to CDW of 1\emph{T}-TaSe$_2$,\cite{Myron-1977} we still consider that it is farfetched to connect the nesting vector with a CDW vector of $\emph{\textbf{q}}_{\textrm{CCDW}}=\frac{3}{13}\emph{\textbf{a}}^*+\frac{1}{13}\emph{\textbf{b}}^*$.

On the other hand, we investigated whether the PLD is coupled with CDW. We have calculated the electron-phonon coupling in the $\emph{\textbf{q}}_z$ = 0 plane for 1\emph{T}-TaSe$_2$, 1\emph{T}-TaSeTe(O), and 1\emph{T}-TaTe$_2$. Figure~\ref{Fig_gamma} shows the calculated phonon linewidth $\gamma$ of the lowest phonon modes in the $\emph{\textbf{q}}_z$ = 0 plane. The phonon linewidth $\gamma$ is defined by
\begin{eqnarray}
\gamma_{\textbf{\emph{q}}\nu}=2\pi\omega_{\textbf{\emph{q}}\nu}\sum_{ij}\int\frac{d^3k}{\Omega_{BZ}}|g_{\textbf{\emph{q}}\nu}(\textbf{\emph{k}},i,j)|^2\nonumber\\
\times\delta(e_{\textbf{\emph{q}},i}-e_\textbf{\emph{F}})\delta(e_{\textbf{\emph{k}}+\textbf{\emph{q}},j}-e_\textbf{\emph{F}}),
\label{Eq_gammar}
\end{eqnarray}
in which the electron-phonon coefficients $g_{\textbf{\emph{q}}\nu}(\textbf{\emph{k}},i,j)$ are defined as,
\begin{equation}
g_{\textbf{\emph{q}}\nu}(\textbf{\emph{k}},i,j)=(\frac{\hbar}{2M\omega_{\textbf{\emph{q}}\nu}})^{1/2}\langle\psi_{i,\textbf{\emph{k}}}|\frac{dV_{\textrm{SCF}}}{d\hat{u}_{\textbf{\emph{q}}\nu}}\cdot\hat{\epsilon}_{\textbf{\emph{q}}\nu}|\psi_{j,\textbf{\emph{k}}+\textbf{\emph{q}}}\rangle.
\label{Eq_chi}
\end{equation}
According to this definition, $\gamma$, which reflects the electron-phonon coupling contribution, is a quantity that does not depend on real or imaginary nature of the phonon frequency.\cite{Calandra-NbS2} Although the calculation with $16 \times 16 \times 1$ \textbf{\emph{q}}-points is not enough to deduce the accurate vector with maximum $\gamma$, it can still qualitatively reflect the role of electron-phonon coupling. In the instability area, the $\gamma$ of the lowest mode is hundreds times larger than those of higher modes, proving the connection between electron-phonon coupling and CDW. For 1\emph{T}-TaSe$_2$, the biggest $\gamma$ ($\sim$23.11 GHz) is found near the place where $\chi^{\prime\prime}$ shows the maximum, which is understandable since Fermi surface nesting can enhance $\gamma$ according to Eqs.(~\ref{Eq_gammar}) and (~\ref{Eq_chi}). The second largest $\gamma$ ($\sim$22.18 GHz) is found in the place very near the reported $\emph{\textbf{q}}_{\textrm{CDW}}=\frac{3}{13}\emph{\textbf{a}}^*\textrm{+}\frac{1}{13}\emph{\textbf{b}}^*$. Therefore, if we neglect the enhancement of nesting, one can find the area of \textbf{\emph{q}}-points with large $\gamma$ is centered at the reported $\emph{\textbf{q}}_{\textrm{CDW}}$. For 1\emph{T}-TaTe$_2$ and 1\emph{T}-TaSeTe(O), the large $\gamma$ area is largely broadened and expands to $\Gamma$K, which is coincide with the phonon instability area shown in Fig.~\ref{Fig_phonon}. Meanwhile, $\chi^\prime$ and $\chi^{\prime\prime}$ show small values in the place between $\Gamma$ and K points. Therefore, we can conclude that the \textbf{\emph{q}}-dependent electron-phonon coupling induced PLD, instead of Fermi surface nesting, is responsible for CDW in 1\emph{T}-TaSe$_{2-x}$Te$_x$ system.

Moreover, if we consider the PLD is the origin of CDW, the disappearance of CDW in the phase diagram is understandable. From Table~\ref{Structure_LDA}, one can notice that the optimized \emph{z}-coordinates of X atoms in pristine 1\emph{T}-TaX$_2$ (X = Se, Te) is about $\pm0.27$. However, for 1\emph{T}-TaSeTe(O), the \emph{z}-coordinates of X atoms change to $z_{\textrm{Se}}=0.249$ and $z_{\textrm{Te}}=-0.296$, which means the TaX$_6$ octahedra are largely distorted. When the Se and Te atoms are randomly mixed, disordered distortions of TaX$_6$ octahedra can be expected in the crystal, leading to the puckered Ta-Ta layers. Obviously it is not compatible with two-dimensional PLD. That might be the reason why disorder completely suppresses CDW in 1\emph{T}-TaSe$_{2-x}$Te$_x$ system.

To understand the difference between 1\emph{T}-TaSe$_{2-x}$Te$_x$ and 1\emph{T}-TaS$_{2-x}$Se$_x$, a similar hypothetic 1\emph{T}-TaSSe(O) was also designed and calculated. Figure~\ref{Fig_TaSSe_phonon} compares the phonon dispersion of 1\emph{T}-TaSeTe(O) and that of 1\emph{T}-TaSSe(O). The instability of 1\emph{T}-TaSSe(O) is only found between $\Gamma$ and M points. One can expect that a small area of instability should exist in 1\emph{T}-TaSSe(O), just like that in 1\emph{T}-TaSe$_2$ (Fig.~\ref{Fig_phonon}(b)). Such similarity might indicate that the ordered structure of TaSSe can potentially be prepared in 1\emph{T} structure. On the other hand, the instability of 1\emph{T}-TaSeTe(O) is broadened and expands to the $\Gamma$K line, just like the case in 1\emph{T}-TaTe$_2$. The large area of instability leads to a very stable single-\textbf{\emph{q}} CDW-type distortion (monoclinic structure) in TaTe$_2$, so that the ideal 1\emph{T}-structure has never been observed in TaTe$_2$. Similarly, if TaSeTe has the ordered structure, it would show monoclinic structure as well. However, our experimentally prepared TaSeTe clearly shows 1\emph{T} structure. According to such analysis, one can conclude the ordered structure of 1\emph{T}-TaSSe(O) should be more stable than that of 1\emph{T}-TaSeTe(O). In other words, the disorder in 1\emph{T}-TaSeTe is much stronger than that in 1\emph{T}-TaSSe. It could explain that in 1\emph{T}-TaS$_{2-x}$Se$_x$ ($0 \leq x \leq 2$) CDW exists with all \emph{x}, while in 1\emph{T}-TaSe$_{2-x}$Te$_x$ ($0 \leq x \leq 2$) the CDW or CDW-type distortion is suppressed as $0.5 < x < 1.5$. Very recently, the ordered structure of 1\emph{T}-TaSSe(O)\cite{Ang-Natcommun} and the disorder in 1\emph{T}-TaSeTe\cite{Huixia-PNAS} have been experimentally suggested, which support our deduction.

The PLD mechanism might help to understand the appearance of superconductivity in this system. The electron-phonon coupling strength for each mode ($\lambda_{\textbf{\emph{q}}\nu}$) is defined as,
\begin{equation}
\lambda_{\textbf{\emph{q}}\nu}=\frac{\gamma_{\textbf{\emph{q}}\nu}}{\pi\hbar N(e_\textbf{\emph{F}})\omega^2_{\textbf{\emph{q}}\nu}}.
\label{Eq_lambda}
\end{equation}
An imaginary frequency $\omega$ of the phonon mode indicates the phase instability (in our case it indicates the CDW distortion). When the CDW is suppressed, the stabilizing of 1\emph{T} structure will make the imaginary frequency $\omega$ around $\textbf{\emph{q}}_{\textrm{CDW}}$ become a small real value, just like the cases of 1\emph{T}-TaS$_2$ and 1\emph{T}-TaSe$_2$ under pressure.\cite{AmyLiu-1T-TaS2, AmyLiu-1T-TaSe2} The large $\gamma$ and small real $\omega$ in Eq.(~\ref{Eq_lambda}) can lead to a large electron-phonon coupling constant, which might be the reason why superconductivity emerges when CDW is suppressed. Further measurements are needed to verify the above arguments.

\section{Conclusion}
In order to connect the CDW of 1\emph{T}-TaSe$_2$ and the CDW-type distortion of 1\emph{T}-TaTe$_2$, we prepared a series of 1\emph{T}-TaSe$_{2-x}$Te$_x$ ($0 \leq x \leq 2$) single crystals and summarized an overall electronic phase diagram through the transport measurements, in which a dome-like superconducting region is observed. The CDW disappears in 1\emph{T}-TaSe$_{2-x}$Te$_x$ as $0.5 < x < 1.5$, which is unexpected since the similar isovalent doping in 1\emph{T}-TaS$_{2-x}$Se$_x$ does not seem to completely suppress CDW.

In order to understand the experimental results, we performed DFT calculations on 1\emph{T}-TaSe$_2$, 1\emph{T}-TaTe$_2$, and the hypothetic ordered 1\emph{T}-TaSeTe. 1\emph{T}-TaSe$_2$ and 1\emph{T}-TaTe$_2$ show similar phonon dispersions and instabilities, indicating the distortions of the two end members originate from the same mechanism. Similar instability is also found in the hypothetic ordered 1\emph{T}-TaSeTe while CDW disappears in experimentally prepared 1\emph{T}-TaSeTe, implying CDW in real 1\emph{T}-TaSe$_{2-x}$Te$_x$ is suppressed by disorder. Based on the generated electron susceptibility calculations, the formation and suppression of CDW in 1\emph{T}-TaSe$_{2-x}$Te$_x$ are found to be independent of Fermi surface nesting. Through analysis of the optimized structures of 1\emph{T}-TaSe$_2$, 1\emph{T}-TaTe$_2$, and 1\emph{T}-TaSeTe(O), we found that the Te doping can largely distort the TaX$_6$ (X = Se, Te) octahedra. The disordered distribution of those distorted octahedra will pucker Ta-Ta layers, which is not compatible with the two-dimensional PLD. That might be why CDW is suppressed in 1\emph{T}-TaSe$_{2-x}$Te$_x$ system. The interplay of CDW and superconductivity is also discussed. Our results offer an indirect evidence that the PLD, which can be influenced by strong disorder, is the origin of CDW in the system.

\begin{acknowledgments}
This work was supported by the National Key Basic Research under Contract No. 2011CBA00111, the National Nature Science Foundation of China under Contract Nos. 11304320, 11404342, 11204312 and 11274311, the Joint Funds of the National Natural Science Foundation of China and the Chinese Academy of Sciences¡¯ Large-scale Scientific Facility (Grand No. U1232139), Anhui Provincial Natural Science Foundation under Contract No. 1408085MA11, and the Director¡¯s Fund under Contract No.YZJJ201311 of Hefei Institutes of Physical Science, Chinese Academy of Sciences.

\end{acknowledgments}

\end{document}